\documentclass[11pt,fleqn,a4paper]{article}
\pdfoutput=1 % if your are submitting a pdflatex (i.e. if you have
\usepackage[colorlinks=true, linkcolor=black!50!blue, urlcolor=blue, citecolor=blue, anchorcolor=blue]{hyperref}
\usepackage[font=small,labelfont=bf,margin=0mm,labelsep=period,tableposition=top]{caption}
\usepackage[a4paper,top=3cm,bottom=2.5cm,left=2.5cm,right=2.5cm,bindingoffset=0mm]{geometry}
\usepackage[T1]{fontenc} % if needed

\usepackage[latin1]{inputenc}
\usepackage[english]{babel}

\usepackage{pifont}
\pdfoutput=1
\usepackage{graphicx,mathtools}
\usepackage{epsfig,cite}
\usepackage{amssymb}
\usepackage{amsmath}
\usepackage{color}
\usepackage{amstext,alltt,setspace}
\usepackage{amsbsy}
\usepackage{array}
\usepackage{booktabs, makecell}
\usepackage{hyperref}
\usepackage{slashed}
\usepackage{url}
\usepackage{geometry}
\usepackage{amssymb,amsthm,cancel,hyperref,graphicx,xcolor}
\usepackage{picinpar,graphicx,xy}
\usepackage[subnum]{cases}
\usepackage{booktabs}
\usepackage{mathrsfs}
\usepackage{amsfonts}
\usepackage{latexsym}
\usepackage{booktabs,graphicx,hyperref,epsfig,multirow}
\usepackage{bm}
\usepackage{comment}
\def\bea{\begin{eqnarray}}
\def\eea{\end{eqnarray}}
\def\beq{\begin{equation}}
\def\eeq{\end{equation}}
\def\ba{\begin{eqnarray}}
\def\ea{\end{eqnarray}}
\def\be{\begin{equation}}
\def\ee{\end{equation}}
\definecolor{darkgreen}{HTML}{008000}
\newcommand{\sss}{\scriptscriptstyle\rm}
\newcommand{\muf}{\mu_{\rm\sss F}}
\newcommand{\mur}{\mu_{\rm\sss R}}

\newcommand{\Et}{E_{\sss T}}

\newcommand{\as}{\alpha_s}

\def\({\left(}
\def\){\right)}
\def\[{\left[}
\def\]{\right]}
\def    \hepph  #1 {{\tt hep-ph/#1}}
\def    \hepex  #1 {{\tt hep-ex/#1}}
\long\def\symbolfootnote[#1]#2{\begingroup%
\def\thefootnote{\fnsymbol{footnote}}\footnote[#1]{#2}\endgroup}

\def\lapprox{\lower .7ex\hbox{$\;\stackrel{\textstyle <}{\sim}\;$}}
\def\gapprox{\lower .7ex\hbox{$\;\stackrel{\textstyle >}{\sim}\;$}}

\renewcommand{\(}{\left(}
\renewcommand{\)}{\right)}

\newcommand{\pt}{p_{\sss T}}

\graphicspath{{Images/}}
\begin{document}
\begin{flushleft}
%%%%%%%%%%%%%%%%%%%%%%%%%%%%%%%%%%%%%%%%%%%%%%%%
\begin{figure}[h]
\includegraphics[width=.2\textwidth]{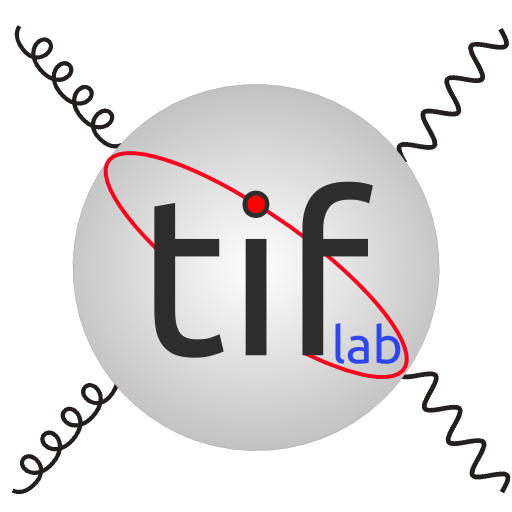}
\end{figure}
%%%%%%%%%%%%%%%%%%%%%%%%%%%%%%%%%%%%%%%%%%%%%%%%%
\end{flushleft}
\vspace{-5.0cm}
\begin{flushright}
TIF-UNIMI-2021-08
\end{flushright}

\vspace{2.0cm}

\begin{center}
{\Large \bf Threshold resummation of transverse momentum distributions
  beyond next-to-leading log}\\
\end{center}

\vspace{1.3cm}

\begin{center}
Stefano Forte$^1$, Giovanni Ridolfi$^2$, Simone Rota$^1$ \\
\vspace{.3cm}
{\it
{}$^1$Tif Lab, Dipartimento di Fisica, Universit\`a di Milano and\\
INFN, Sezione di Milano, Via Celoria 16, I-20133 Milano, Italy\\
{}$^2$Dipartimento di Fisica, Universit\`a di Genova and\\
INFN, Sezione di Genova, Via Dodecaneso 33, I-16146 Genova, Italy.\\
}

{\bf \large Abstract}
\end{center}
We derive a general expression for the threshold resummation of
transverse momentum distributions for processes with a colorless final
state, by suitably generalizing the renormalization-group based
approach to threshold resummation previously pursued by two of us. The
ensuing expression holds to all logarithmic orders, and
it can be used to extend available results in
the literature, which only hold up to the next-to-leading log (NLL)
level. We check agreement of our result with the existing NLL result,
as well as against the
known fixed  next-to-leading order results for the Higgs transverse momentum
distribution in gluon fusion, and we provide explicit expressions at
the next-to-next-to-leading log level.
\clearpage
\tableofcontents
\clearpage

\section{Introduction}
\label{sec:introduction}
Sudakov resummation is routinely used to improve the domain
of validity and the perturbative convergence of fixed-order calculations of differential observables at the LHC. For
example, the transverse momentum distribution of
Higgs and gauge 
bosons produced in hadronic collisions
has been recently determined at matched NNLO+N$^3$LL, i.e., 
fixed next-to-next-to-leading-order and resummed
next$^3$-leading-logarithmic order in transverse-momentum
logarithms~\cite{Bizon:2018foh,Bizon:2019zgf,Ebert:2020dfc,Billis:2021ecs}. However,  threshold
resummation is typically not included in the computation of
differential observables,
despite the fact that recent work has shown that combining threshold
on top of   transverse momentum resummation has a significant impact
on perturbative
convergence~\cite{Muselli:2017bad,Rabemananjara:2020rvw}. This is
especially surprising in view of the common use of threshold
resummation to improve the perturbative convergence of total
cross-sections~\cite{Bonvini:2016frm}.
In fact, it turns out that threshold resummation for transverse
momentum distributions is currently only available in direct QCD up to
next-to leading log (NLL)~\cite{deFlorian:2005fzc}, with NNLL results
available only in SCET~\cite{Huang:2014mca,Becher:2014tsa}

As pointed out in Ref.~\cite{deFlorian:2005fzc}, threshold resummation
for transverse momentum distributions is structurally similar to --- and can
be viewed as a generalization of --- the soft resummation of the total
cross-section for prompt-photon production~\cite{Catani:1998tm} to
which it may be related using the general resummation formalism for
multiparton processes of Ref.~\cite{Bonciani:2003nt}. A general
approach to soft resummation based on renormalization group (RG) arguments
was proposed by some of us long ago~\cite{Forte:2002ni}, based on the
earlier approach of Ref.~\cite{Contopanagos:1996nh}, and shortly
  thereafter extended to include the case of prompt-photon
  production~\cite{Bolzoni:2005xn}. This approach has the advantage of
  generality: it
  provides the form of resummed results to any logarithmic order,
  though it does not allow for the computation of the coefficients
  that determine the resummed expressions explicitly, which must be
  found by matching to fixed-order expressions --- the fixed
  N$^k$LO fully determines the N$^k$LL resummed result.

Here, we extend the approach of Ref.~\cite{Forte:2002ni} to transverse
momentum distributions, by viewing them as a generalization of the
case of prompt-photon production~\cite{Bolzoni:2005xn}. We arrive at a
form of the soft resummation which is in fact somewhat more compact
than that of Ref.~\cite{deFlorian:2005fzc}, with which we prove
agreement at NLL. We work out  the general form of the
resummed expressions up to NNLL. We check the agreement with fixed NLO
result for Higgs production in gluon fusion~\cite{Glosser:2002gm}, which
verifies  the correctness of the resummed result up to  NLL. This
comparison was  already performed 
numerically in Ref.~\cite{deFlorian:2005fzc}; here  we show that the
resummation correctly predicts the logarithmically enhanced terms in
the fixed-order result. A check of the NNLL result would require 
comparing the NNLO fixed order. While for Higgs production
in gluon fusion this is
principle available~\cite{Chen:2016zka,Lindert:2018iug}, in practice
the explicit analytic expression which would be needed in order to read
off the relevant coefficient is not readily obtained.

The generalization of the RG argument of
Ref.~\cite{Bolzoni:2005xn} to transverse-momentum distributions
requires an analysis of the phase space structure for real emission in
the soft limit that will be presented in
Sect.~\ref{sec:phasespace}. This will be used in Sect.~\ref{sec:rg} to derive the general
form of resummed results, that in Sect.~\ref{sec:resexp} will be
compared to available  results.

\section{Phase space factorization}
\label{sec:phasespace}
We consider  the transverse momentum distribution
$\frac{d\sigma}{d\pt^2}$ for the process
\beq
\label{HHg}
H_1(P_1)+H_2(P_2)\rightarrow H+X,
\eeq
where $H$ is a final-state object (particle or
system), whose invariant mass will be denoted by $m$, and  $\pt$ is the transverse momentum of $H$ with respect to
the direction of the colliding hadrons $H_1$ and $H_2$.

\subsection{Kinematics}
\label{sec:kin}

The transverse momentum distribution is characterized by two scales,
which can be constructed out of the the invariant mass $m$ and transverse momentum $\pt$, and a scaling
variable $\tau$. Factorization follows if the scaling variable is chosen as
\begin{equation}
  \label{eq:scalvara}
  \tau\equiv\frac{(\Et+\pt)^2}{s},
\end{equation}
where $s=(P_1+P_2)^2$
    and we denote with $\pt$ the modulus of the transverse momentum vector and  with $\Et$ the transverse energy
\begin{align}
\Et= \sqrt{m^2+ \pt^2};\qquad \pt=|{\vec p}_{\sss T}|.
     \label{eq:etpt}
\end{align}
This suggests the choice of 
    \begin{align}\label{q2}
      Q=\sqrt{m^2+ \pt^2}+\sqrt{\pt^2}=\Et+\pt
    \end{align}
    as one of the two scales of the process, so that
\begin{equation}
  \label{eq:scalvar}
  \tau=\frac{Q^2}{s},
\end{equation}
As we shall see below, a natural choice for the other independent
scale is $Q\pt$: indeed this was denoted as $Q^2$ in
Ref.~\cite{deFlorian:2005fzc}, and we shall therefore consider
henceforth the cross-section as a function of $Q^2$ and $Q\pt$.

The scale $Q$ Eq.~(\ref{q2})
is the threshold energy, i.e.\ the minimum energy needed to produce a
system with invariant mass $m$ and transverse momentum $\pt$. This
ensures factorization in that,
with this choice, the kinematic boundary for the scaling
variable $\tau$ at fixed $\pt$ is $\pt$-independent:
\begin{equation}\label{eq:kinboud}
     0\le \tau\le 1
\end{equation}
so 
\begin{align}
\label{eq:conv2}
\frac{d\sigma}{d\pt^2}\(\tau,Q^2,Q\pt\)=\tau\sum_{i,j}\int_{\tau}^1 \frac{dx}{x}\, \mathcal{L}_{ij}\(\frac{\tau}{x}\)\frac{1}{x}\frac{d\hat{\sigma}_{ij}}{d\pt^2}\(x,Q^2,Q\pt\),
\end{align}
where 
\be
x=\frac{Q^2}{\hat s}=\frac{(\sqrt{m^2+\pt^2}+\pt)^2}{\hat s},
\label{eq:partx}
\ee
$\hat s$ is the center-of-mass partonic energy squared, and
the parton luminosity is defined in the usual way as
\beq
\mathcal{L}_{ij}(x)=\int_{x}^1 \frac{dy}{y} f_i\(y\)f_j\(\frac{x}{y}\).
\eeq
To keep notations simple, the dependence of the quantities involved on the strong coupling $\as(\mur^2)$ and on the factorization scale $\muf$ are omitted in this section.

Because of the $\pt$ independence of the limits of integration,
Eq.~(\ref{eq:conv2}) factorizes upon taking a Mellin transform with
respect to the scaling variable $\tau$. Namely, defining
\begin{align}
\label{eq:defMellinhadro}
\frac{d\sigma}{d\pt^2}\(N,Q^2,Q\pt\)&=\int_0^1 d\tau\,\tau\,^{N-1}
\frac{d\sigma}{d\pt^2}\(\tau,Q^2,Q\pt\);\\
\label{eq:defMellin}
\frac{d\hat{\sigma}_{ij}}{d \pt^2}\(N, Q^2,Q\pt\)&=\int_0^1 dx\, x^{N-1} \frac{d\hat{\sigma}_{ij}}{d \pt^2}\(x, Q^2,Q\pt\),
\end{align}
where with a slight abuse of notation we are using the same symbol for
the cross section and its Mellin transform,
we get
\beq
\label{eq:conv3}
\frac{d\sigma}{d\pt^2}\(N, \pt^2,m^2\)=\sum_{ij} \mathcal{L}_{ij}\(N+1,\muf^2\) \frac{d\hat{\sigma}_{ij}}{d \pt^2}\(N,Q^2,Q\pt\).
\eeq
Note that  because the scale $Q$
Eq.~(\ref{q2}), and consequently the scaling variable
Eq.~(\ref{eq:scalvar}) depend on $\pt$, they necessarily
differ from the scale and scaling variable used for the total cross-section.

It is also useful to introduce the variable
\begin{equation}\label{eq:qbar}
       \bar Q\equiv \Et-\pt,
\end{equation}
so that
 \begin{equation}\label{eq:ptq}
 \pt=\frac{1}{2}\left(Q-\bar Q\right)
     \end{equation}
and
\begin{equation}\label{eq:Etq}
\Et=\frac{1}{2}\left(Q+\bar Q\right).
\end{equation}
The definition Eq.~(\ref{eq:qbar}) implies that
     \begin{equation}
       Q\bar Q= \Et^2- \pt^2=m^2
     \end{equation}   
or
\begin{equation}\label{eq:qqbar}
\bar Q=\frac{m^2}{Q}.
\end{equation}
It follows in particular that
substituting  the expression Eq.~(\ref{eq:qqbar}) in the expression
Eq.~(\ref{eq:ptq}) of $ \pt$  we get
     \begin{equation} 
       Q \pt= \frac{Q^2-m^2}{2 },
       \label{eq:ptQ}
     \end{equation}
     which shows that  any pair of variables among $ Q\pt$,
     $Q^2$, $m^2$ can be chosen
     as independent kinematic variables, along with the dimensionless
ratio $\tau$ Eq.~(\ref{eq:scalvar}).
With any such choice  there are two scales and a
scaling variable, which can be varied independently without
conflicting with factorization, i.e. in such a way that at the
factorized level the parton luminosity only depends on $\tau$ and a scale.

\subsection{Phase space}
\label{sec:phsp}

We now consider the phase space measure for the process Eq.~(\ref{HHg}) in the
soft limit. We will follow the treatment of
Ref.~\cite{Bolzoni:2005xn}, which in turn exploits the general
formalism for dealing with an $n$-body phase-space discussed in the
appendix of Ref.~\cite{Forte:2002ni}, recently generalized,
systematized, and supplemented with a graphical representation in Ref.~\cite{Jing:2020tth}.

Specifically,
we consider a contribution to the transverse momentum distribution
in which there are $k+1$ massless partons in the final state, with momenta
$k_1,\ldots,k_{k+1}$, and
\beq
\label{consg}
p_1+p_2=p_H+k_1+\ldots +k_{k+1}.
\eeq
We are interested in the threshold limit. For a $\pt$ distribution
Eq.~(\ref{HHg}), 
there must be at least a non-soft final state parton in $X$ that recoils
against  $H$. Hence, the threshold limit can be approached when all
other partons are either collinear to this non-soft parton, or soft.
We then assume that  momenta
$k_i,i=1,\ldots,n; n\le k$ are soft, while
momenta $k_i,i>n$ are non-soft. For the sake of simplicity, 
we relabel non-soft momenta as
\beq
k'_j =k_{n+j};\qquad 1\leq j\leq m+1;\qquad m=k-n.
\eeq
The generic kinematic configuration in the soft limit is then
\bea
&&k_i=0\qquad 1\leq i\leq n
\nonumber\\
&&\theta_{ij}=0;\qquad \sum_{j=1}^{m+1} k^{'0}_j= \pt\qquad
\qquad 1\leq i,j\leq m+1
\eea
($\theta_{ij}$ being the angle formed by $\vec k_i$ and $\vec k_j$)
for all $n$ between $1$ and $k$, namely, the configuration where at
least one momentum is not soft, and the remaining momenta are either
collinear to it, or soft.

With this labeling of the momenta, the phase space can be written as
\bea
\label{genidg}
&&d\phi_{n+m+2}(p_1+p_2;p_H,k_1,\ldots,k_n,k'_1,\ldots,k'_{m+1})
\\
&=&\int
\frac{dq^2}{2\pi}d\phi_{n+1}(p_1+p_2;q,k_1,\ldots,k_n)
\int\frac{d{k'}^2}{2\pi}\,d\phi_2(q;p_H,k')\,
d\phi_{m+1}(k';k'_1,\dots,k'_{m+1}).
\nonumber
\eea
Here:
\begin{itemize}
  \item $d\phi_{n+1}$ is the inclusive phase space for
a production process with two incoming partons with momenta $p_1$, $p_2$,
of a massive object with mass $q^2$, plus $n$ partons with momenta
$k_i$ (to be taken as soft).
\item $d\phi_2$ is the phase space for production, from an incoming
  momentum $q$,  of a massive final
  state with mass $m$, whose momentum $p_H$ will be assumed to have a
  fixed transverse momentum $ \pt$, and a system with momentum $k'$
  recoiling against it. Note that in the soft limit in which all momenta
  $k_i$ vanish, $q$ reduces to $p_1+p_2$, so $q^2=s$ .
\item  $d\phi_{m+1}$ is the phase space for the
  production, from incoming momentum $k'$, of a final-state system
  containing $m+1$ partons with momenta $k'_i$. Note that in the soft
  limit, in which all momenta $k_i$ vanish, and all momenta $k'_i$ are
  collinear to $k'$, momenta $p_H$ and $k'$ are back-to-back. Note
  also that in the small $ \pt$ limit all momenta $k'$ are also
  collinear to the incoming parton's direction (but not soft).
\end{itemize}

We first work out the two-body phase space  $d\phi_2$. In
$4-2\epsilon$ dimensions and in the rest frame of $q$ we have
\begin{align}
d\phi_2(q;p_H,k')
&=\frac{d^{d-1}k'}{(2\pi)^{d-1}2k^{'0}}\,
\frac{d^{d-1}p_H}{(2\pi)^{d-1}2p_H^0}\,
(2\pi)^d\delta^{(d)}(q-k'-p_H)
\nonumber\\
&=\frac{(4\pi)^\epsilon}{16\pi\Gamma(1-\epsilon)}
 \pt^{-2\epsilon}
\frac{d \pt^2dp_z}{k_0'p_H^0}\delta(p_H^0+k_0'-\sqrt{q^2})
\label{eq:dp2g}.
\end{align}
The delta can be used to perform the integration in $p_z$. In the rest frame of $q$ we find
\be
\delta(p_H^0+k_0'-\sqrt{q^2})=\frac{\delta(p_z-\bar p_z)}{|J(\bar p_z)|}
\ee
where
\be
\bar p_z^2=\frac{\lambda(m^2,q^2,{k'}^2)}{4q^2}- \pt^2;\qquad
\lambda(x,y,z)=x^2+y^2+z^2-2xy-2xz-2yz
\label{pz}
\ee
and
\be
|J(\bar p_z)|=\bar p_z\(\frac{1}{p^0_H}+\frac{1}{k_0'}\)=\frac{\bar p_z\sqrt{q^2}}{p^0_Hk_0'}.
\ee
Hence
 \be
d\phi_2(q;p_H,k')=\frac{(4\pi)^\epsilon}{16\pi\Gamma(1-\epsilon)}
 \frac{\pt^{-2\epsilon}}{\bar p_z\sqrt{q^2}}d \pt^2.
\ee

Next, we work out the kinematic limits for the integration variables
$q^2$ and ${k'}^2$. 
  We have
  \begin{align}
    p_1+p_2&=q+\sum_{i=1}^nk_i\label{eq:qoutg}\\
    q&=p_H+k'.\label{eq:qing}
  \end{align}
We start with the bounds for $q^2$.
  Equation~(\ref{eq:qoutg}) implies $q^2\le \hat s=(p_1+p_2)^2$. Eq.~(\ref{eq:qing})
implies, in the rest frame of  $q$, which in the soft limit coincides
with the center-of-mass frame, 
\begin{equation}
  q^2=\left(\sqrt{m^2+ \pt^2+p_z^2}+\sqrt{{k'}^2+ \pt^2+p_z^2}\right)^2\label{eq:qming}
\end{equation}
so the minimum value of $q^2$ is attained when ${k'}^2=p_z=0$, and it is
equal to $q_{\rm min}^2=Q^2$. Hence we conclude that
\begin{equation}
  Q^2\le q^2\le \hat s.
  \label{eq:qlimsg}
\end{equation}
We consider next the bounds on ${k'}^2$.
Equation~(\ref{eq:qing}) implies
\begin{equation}
  {k'}^2=q^2+m^2-2\sqrt{q^2}\sqrt{m^2+ \pt^2+p_z^2}
  \label{eq:ktprmg}
  \end{equation}
so the maximum value of $k'$ is attained when $p_z=0$:
\begin{align}
   {k'}_{\rm max}^2&=q^2+m^2-2\sqrt{q^2}\Et\nonumber\\
  &=q^2+m^2-\sqrt{q^2} \left(Q+\frac{m^2}{Q}\right),
  \label{eq:kpmax1g}
\end{align}
where in the last step we have used
Eqs.~(\ref{eq:Etq},\ref{eq:qqbar}). 

We can finally write the full phase space as follows:
\begin{align}
&d\phi_{n+m+2}(p_1+p_2;p_H,k_1,\ldots,k_n,k'_1,\ldots,k'_{m+1})
= \frac{(4\pi)^\epsilon}{64\pi^3\Gamma(1-\epsilon)} \pt^{-2\epsilon}d \pt^2
\nonumber\\
&\qquad\times\int_{Q^2}^s \frac{dq^2}{\sqrt{q^2}}
\int_0^{{k'}^2_{\rm max}}
\frac{d{k'}^2}{\bar p_z}
d\phi_{n+1}(p_1+p_2;q,k_1,\ldots,k_n) 
d\phi_{m+1}(k';k'_1,\dots,k'_{m+1}).
\end{align}

We now consider the soft limit, $x\to 1$.
We introduce a dimensionless parameter $0\le u\le 1$
interpolating between the two extremes for $q^2$. We have
\begin{equation}
  q^2=
  Q^2+u(\hat s-Q^2)=Q^2\left(1+u\frac{1-x}{x}\right)\label{eq:dykin}
\end{equation}
which shows that $q^2\to Q^2$ in the soft limit.

Using Eq.~(\ref{eq:dykin}) to express
$q^2$ in terms of the independent kinematic variables we get
\begin{align}
{k'}_{\rm max}^2&=Q^2\left(1+u\frac{1-x}{x}\right)+m^2
-(Q^2+m^2)\sqrt{1+u\frac{1-x}{x}}
\nonumber\\
&=\frac{1}{2} (Q^2-m^2) u (1-x)+O\((1-x)^2\)
\nonumber\\
&=Q  \pt u (1-x)+O\((1-x)^2\),
\label{eq:kpmax2g}
\end{align}
where we have used Eq.~(\ref{eq:ptQ}) in the last step.
Introducing a further dimensionless parameter $0\le v\le1$ in order to interpolate
between the two extremes we have
\begin{equation}
 {k'}^2= uv Q \pt(1-x).
  \label{eq:kpparg}
\end{equation}
Hence in the soft limit $ {k'}^2\to 0$.

Finally, in this limit,
\be
q^2\bar p_z^2= \pt \Et Q^2(1-x)u(1-v)\[1+O(1-x)\]
\ee
and therefore
\begin{align}
&d\phi_{n+m+2}(p_1+p_2;p_H,k_1,\ldots,k_n,k'_1,\ldots,k'_{m+1})
\nonumber\\
&= \frac{(4\pi)^\epsilon}{64\pi^3\Gamma(1-\epsilon)} \pt^{-2\epsilon}d \pt^2
Q^2(1-x)^{3/2}\sqrt{\frac{\pt}{\Et}}
\int_0^1du\int_0^1dv\,\sqrt{\frac{u}{1-v}}
\nonumber\\
&
d\phi_{n+1}(p_1+p_2;q,k_1,\ldots,k_n) 
d\phi_{m+1}(k';k'_1,\dots,k'_{m+1}).
\end{align}
The full phase space thus factors into the product of two phase
spaces, related by the $u$ and $v$ integrations, with $ \pt$
kept fixed.

Phase space
$d\phi_{n+1}(p_1+p_2;q,k_1,\ldots,k_n)$ is the same as that for
Drell-Yan or Higgs production, namely for the production of a
colorless final
state of mass $q^2$, as given in Ref.~\cite{Forte:2002ni} (see in
particular Eq.~(4.30) of that reference).  Note that it does not
depend on $ \pt$.
In the soft limit, this  phase space can be written in terms of a dimensionless
integration measure, with all the dimensional dependence contained in
a prefactor,  given by a power of
\be\label{eq:softscale}
\frac{(\hat s-q^2)^2}{q^2}\propto Q^2(1-x)^2\equiv\Lambda^2_{\rm DY}.
\ee

Phase space $d\phi_{m+1}(k';k'_1,\dots,k'_{m+1})$
can be viewed as a phase space with the same structure of
deep-inelastic scattering, namely incoming momentum ${k'}^2$, with the
variable $k'$ now integrated over, and vanishing in the soft
limit, again as  given in Ref.~\cite{Forte:2002ni} (see in
particular Eq.~(4.17) of that reference). This too can be written in
terms of a dimensionless integration measure, with now the
dimensional dependence contained in a power of 
\be\label{eq:collscale}
 {k'}^2\propto Q \pt(1-x)\equiv \Lambda^2_{\rm DIS}.
\ee

In summary, in the soft limit the phase space for transverse momentum distributions
factorizes completely into a Drell-Yan-like phase space, related to
soft emission, that only
depends on the dimensional scale $\Lambda^2_{\rm DY}$ Eq.~(\ref{eq:softscale}), and a DIS-like
phase space, related to collinear emission, that only
depends on the dimensional scale $\Lambda^2_{\rm DIS}$
Eq.~(\ref{eq:collscale}). Note that what determines the scale is not
whether emission is from incoming or outgoing legs, but rather,
whether the emission can contribute in the soft limit because it is
soft, or because it is collinear to the fixed $ \pt$ parton that
recoils against the fixed-$ \pt$ final state $H$.

\section{Resummation}
\label{sec:rg}

The resummation argument is  a rerun of that of
Ref.~\cite{Bolzoni:2005xn}, which generalizes to prompt-photon
production 
the resummation approach developed and discussed in
Refs.~\cite{Forte:2002ni,Contopanagos:1996nh}, specifically for
deep-inelastic scattering (DIS) and Drell-Yan (DY) production. The basic
underlying idea remains the same, but in processes like DIS and DY
one single soft scale is resummed, while in prompt photon production,
as well as in the case of transverse momentum distributions discussed
here, two different soft scales are simultaneously resummed. We first
briefly 
summarize the  argument of
Refs.~\cite{Contopanagos:1996nh,Forte:2002ni,Bolzoni:2005xn} in a
somewhat generalized form, and then we use it to obtain a
resummed expression for transverse momentum distributions exploiting
the results presented in the previous section.

\subsection{The renormalization group argument with a single scale}
\label{sec:rgres}

Resummation is most easily expressed for a coefficient function, which
is 
defined factoring out of the cross-section the Born-level expression. Specifically, for a
total cross-section 
\begin{equation}\label{eq:cfdef}
 \hat{\sigma}_{ij}\(N,Q^2,\as(Q^2)\)=C_{ij}(N,Q^2/\mu^2,\as(\mu^2))\sigma^{0}_{ij}\(N,Q^2,\mu^2,\as(\mu^2)\)
\end{equation}
where $\sigma^{0}_{ij}$ is the leading-order expression, and we have
chosen for simplicity  $\mu^2_F=\mu^2_R=\mu^2$. In the soft limit, only diagonal
partonic channels are unsuppressed,  resummation can be performed
independently in the quark singlet and gluon channel and we will
consequently suppress the parton indices $i,j$.

We discuss first the case of a process with a single hard
scale~\cite{Contopanagos:1996nh,Forte:2002ni}, such as DY or DIS.
Resummation is performed  in terms of the physical anomalous dimension
\begin{align}\label{eq:gamp}
\gamma(N, Q^2/\mu^2,\as(\mu^2),\epsilon)=\frac{d}{d\ln Q^2} \ln
C(N, Q^2/\mu^2,\as(\mu^2),\epsilon),
\end{align}
where we have adopted dimensional regularization with $4-2\epsilon$ space-time dimensions.
The coefficient function is multiplicatively renormalized: in a mass-independent subtraction scheme,
\begin{equation}\label{eq:barecf}
 C(N,Q^2/\mu^2,\as(\mu^2),\epsilon)=Z^C(N,\as(\mu^2),\epsilon)C^{(0)}(N,Q^2,\alpha_0,\epsilon),
\end{equation}
where $C^{(0)}$ and $\alpha_0$ are the bare coefficient function and
coupling respectively. 
It follows that the physical anomalous dimension can be equivalently computed from the
bare coefficient function~\cite{Forte:2002ni}:
\begin{align}\label{eq:gampbare}
\gamma(N, Q^2/\mu^2,\as(\mu^2),\epsilon)=-\epsilon \alpha_0\frac{d}{d\ln \alpha_0} \ln
C^{(0)}(N, Q^2,\alpha_0,\epsilon),
\end{align}
where we have used the fact that $C^{(0)}$ can only depend on $Q^2$ and $\alpha_0$ through the dimensionless 
combination $Q^{-2\epsilon}\alpha_0$.
For a  single-scale process  in the soft limit
the dimensional dependence of the phase
space is through a fixed combination of the scale and the scaling
variable
\begin{equation}\label{eq:gen}
\Lambda^2_a(x,\lambda^2)=\lambda^2(1-x)^a,
\end{equation}
where $a=1$ in the case of DIS, and $a=2$ in the case of DY.
This implies~\cite{Forte:2002ni} that the Mellin-space coefficient function
only depends on $N$ through the dimensional variable
\begin{equation}\label{eq:genn}
\bar\Lambda^2_a(N,\lambda^2)=\frac{\lambda^2}{N^a}. 
\end{equation}
Assuming further full factorization of the soft
singularities~\cite{Contopanagos:1996nh},  this in turn implies that the
coefficient function admits a perturbative expansion of the form
\begin{align}
\label{eq:cbarefact}
&  C^{(0)}(N, Q^2,\alpha_0,\epsilon)=C^{(0,c)}(Q^2,\alpha_0,\epsilon) C^{(0,\,l)}(\bar\Lambda_a^2(N,Q^2),\alpha_0,\epsilon)
\\
\label{eq:cbarec}
&\quad C^{(0,\,c)}(Q^2,\alpha_0,\epsilon)=\sum_n 
C_n^{(0,\,c)}(\epsilon) Q^{- 2n\epsilon}\alpha_0^n
\\\label{eq:cbarel}
&\quad C^{(0,\,l)}(\bar\Lambda^2,\alpha_0,\epsilon)=\sum_n 
C_n^{(0,\,l)}(\epsilon) {\bar\Lambda}^{- 2n\epsilon}\alpha_0^n,
\end{align}
where $C^{(0,\,l)}$ collects contributions due to real emission, which
have nontrivial kinematics, $C^{(0,\,c)}$ collects virtual contributions, that
have Born kinematics, and  factorization is the assumption that virtual
(or ``hard'') and real soft-emission contributions fully factorize.

Equation~(\ref{eq:cbarefact}) implies the decomposition of the
physical anomalous dimension
\begin{align}\label{eq:gampbard}
\gamma\(N,\frac{Q^2}{\mu^2}, \as(\mu^2),\epsilon\)=\gamma^{(c)}\(\frac{Q^2}{\mu^2},\as(\mu^2),\epsilon\)+\gamma^{(l)}\(\frac{\bar\Lambda_a^2(N,Q^2)}{\mu^2},\as(\mu^2),\epsilon\),
\end{align}
where
\begin{align}\label{eq:gaml}
\gamma^{(c)}\(\frac{Q^2}{\mu^2},\as(\mu^2),\epsilon\)&=\epsilon \alpha_0\frac{d}{d\ln
  \alpha_0} \ln C^{(0,\,c)}(Q^2,\alpha_0,\epsilon);\\\label{eq:gamc}
\gamma^{(l)}\(\frac{\bar\Lambda_a^2(N,Q^2)}{\mu^2},\as(\mu^2),\epsilon\)&=\epsilon \alpha_0\frac{d}{d\ln
  \alpha_0} \ln C^{(0,\,l)}(\bar\Lambda_a^2(N,Q^2),\alpha_0,\epsilon).
\end{align}
The two contributions  $\gamma^{(c)}$ and
$\gamma^{(l)}$ are not necessarily separately finite, and thus depend
a priori on the scale $\mu$. However, their sum, the
physical anomalous dimension $\gamma$, is finite and
renormalization-group invariant, so
\begin{equation}\label{eq:sudrge}
-\frac{d}{d\ln\mu^2}  \lim_{\epsilon\to0}\gamma^{(l)}\(\frac{\bar\Lambda_a^2}{\mu^2},\as(\mu^2),\epsilon\)
=\frac{d}{d\ln\mu^2}\lim_{\epsilon\to0}\gamma^{(c)}\(\frac{Q^2}{\mu^2},\as(\mu^2),\epsilon\)=\bar g(\as(\mu^2))
\end{equation}
where $\bar g(\as(\mu^2)$ is a perturbative function of $\as$
with finite coefficients:
\begin{equation}\label{eq:gfun}
  \bar g(\as(\mu^2))=\sum_n \bar g_n \as^n(\mu^2).
\end{equation}
Equation~(\ref{eq:sudrge}) can be viewed as a standard
renormalization-group equation for the physical anomalous dimension,
with solution
\begin{equation}\label{eq:gampbardg}
\gamma(N, 1,\as(Q^2),\epsilon)=
\bar g_0(\as(Q^2))+\int_{Q^2}^{\bar\Lambda_a^2(Q^2, N)} \frac{d\mu^2}{\mu^2}
  \bar g(\as(\mu^2)),
\end{equation}
where $\bar g_0(\alpha)$ is an analytic function of its argument.

\subsection{The resummed coefficient function}
\label{sec:rescf}

Using the expression Eq.~(\ref{eq:gampbardg}) of the physical anomalous dimensions
that emerges from the RG argument leads to a resummed expression for
the coefficient function of the form
\begin{align}\label{eq:rescfdydis}
C(N,Q^2/\mu^2,\as(\mu^2))=C^{(c)}\left(\as(Q^2),\frac{Q^2}{\mu^2}\right)\exp\left[
\int_1^{N^a}\frac{dn}{n} \int_{n \mu^2}^{Q^2}
\frac{dk^2}{k^2}\bar g(\as(k^2/n))\right].
\end{align}
Note that both $\bar g$ and $C^{(c)}$ are power series expansions in
$\as$ starting at order one and zero respectively. Including the
first $k$ terms in the perturbative expansion of both of them  leads to resummation
with N$^{k}$LL accuracy.

It is customary to rewrite the  resummed expression
Eq.~(\ref{eq:rescfdydis}) in such a way that its exponent takes the
form of a Mellin transform. This is done by first, 
performing the change of
integration variable $n=(1-x)^{-a}$, with the result
\begin{align}\label{eq:rescfdydisxnew}
  C(N,Q^2/\mu^2,\as(\mu^2))=C^{(c)}\left(\as(Q^2),\frac{Q^2}{\mu^2}\right)\exp\left[
 a \int_0^{1-\frac{1}{N}}\frac{dx}{1-x}\int_{\mu^2}^{Q^2(1-x)^a}\frac{dk^2}{k^2}\,
  \bar g(\as(k^2))\right]
\end{align}
and then using the identity
\be
\int_0^{1-\frac{1}{N}}\frac{dx}{1-x}\ln^p(1-x)
=-\sum_{n=0}^{p}\binom{p}{n}\Delta^{(n)}(1)\int_0^1dx\frac{x^{N-1}-1}{1-x}\ln^{p-n}(1-x)
+\Delta^{(p+1)}(1)+O\(\frac{1}{N}\)
\label{Ip}
\ee
where $\Delta(z)=1/\Gamma(z)$. One thus finds
\begin{align}\label{eq:rescfdydisx}
  C(N,Q^2/\mu^2,\as(\mu^2))=C^{(c)}\left(\as(Q^2),\frac{Q^2}{\mu^2}\right)\exp\left[
 a \int_0^1dx\,\frac{x^{N-1}-1}{1-x}\int_{\mu^2}^{Q^2(1-x)^a}\frac{dk^2}{k^2}\,
  \hat g(\as(k^2))\right],
\end{align}
where the coefficients of the expansion of the function  $\hat g$ up
to any given order are related to those of $\bar g$ by
Eq.~(\ref{Ip}).

It should be noted, however, that
Eq.~(\ref{eq:rescfdydisx}) is ill-defined, because for $x$ close to 1 the $k^2$ integration range includes the Landau singularity 
of $\as(k^2)$. Hence,  Eq.~(\ref{eq:rescfdydisx}) is only meaningful if
  $\hat g$ is expanded in powers of $\as(Q^2)$ and the $x$ integral is
performed order by order  up to a fixed logarithmic accuracy, i.e.,
in practice going back to the form
Eqs.~(\ref{eq:rescfdydis}-\ref{eq:rescfdydisxnew}).

The advantage of the form Eq.~(\ref{eq:rescfdydisx}) of the resummed
expression is that the resummed exponent is then viewed as the Mellin
transform of perturbative contributions that can be related to eikonal
emission~\cite{Moch:2005ba}. In the specific
case of DIS, and DY (or Higgs):
\begin{align}\label{eq:rescfdydisabd}
&C(N,Q^2/\mu^2,\as(\mu^2))=g_0\left(\as(Q^2),\frac{Q^2}{\mu^2}\right)
\exp\Bigg\{  n\int_0^1dx\,\frac{x^{N-1}-1}{1-x} \int_{\mu^2}^{Q^2(1-x)^a}\frac{dk^2}{k^2}A[\as(k^2)]
\nonumber\\
&+\int_0^1dx\, \frac{x^{N-1}-1}{1-x} \[B[\as(Q^2(1-x)^a)]+ D[\as(Q^2(1-x)^a)]\]\Bigg\},
\end{align}
where again the functions $A$, $B$ and $D$ are perturbative power series in
$\as$. 
Note that this form of the resummation can be always written in the
form of Eq.~(\ref{eq:rescfdydisx}), by expressing the $B$ and
$D$ functions as the integral of a perturbative function of
$\as$, and reabsorbing the contribution from the lower extreme of
integration, which depends on $\mu^2$, in a redefinition of the PDF,
i.e., a change of factorization scheme.  Any $O(\as^n)$ contribution
to $B$ and $D$ will lead to $O(\as^{n-1})$ contributions to $\hat
g$, i.e., they start to contribute at the NLL level.

In Eq.~(\ref{eq:rescfdydisabd}) the coefficients in the expansion of $A$ are just the
coefficients of the most singular contribution ${\frac{1}{(1-x)}}_+$ to the quark or gluon
splitting function expanded in powers of $\as$, with $n$ equal to
the
number of initial-state radiating partons (so $n=1$ for DIS and $n=2$ for DY
and Higgs); $B$ is a universal
function that collects contributions from radiation collinear to a
final state parton (so it is present for DIS but not for DY or Higgs);
and $D$ is a process-dependent function that (starting at
$O(\as^2)$, so NNLL) includes
contributions due to soft but large-angle radiation, and can be shown
to vanish to all orders for DIS~\cite{Gardi:2002xm}, while for DY or
Higgs it can be determined by matching to a fixed-order
computation.

The RG argument only determines the generic form of
the dependence of the coefficient function on the kinematic variable
$N$ (or $x$) which is resummed in the soft limit $N\to\infty$
($x\to1$): its content is that the dependence on $N$ (or $x$)  only goes
through  dimensional scales $\bar \Lambda$ or $\Lambda$ of the form
Eqs.~(\ref{eq:gen}-\ref{eq:genn}), and this in turn through $\as$
in the form of the solution to an RG equation.
In the  cases of DIS and DY (or Higgs) production the phase-space
analysis~\cite{Forte:2002ni} respectively leads to the identification
of $\lambda^2=Q^2$ (exchanged gauge boson  virtuality), $a=1$, or $Q^2=M^2$ (Higgs or
gauge boson mass), $a=2$ in Eqs.~(\ref{eq:gen}-\ref{eq:genn}).
However, this argument does not determine
 the form of the coefficients in the expansion of the $\bar
g_i$ functions, that are fixed, at next$^k$-to-leading logarithmic
order, by comparing to a fixed next$^k$-to-leading order result. 
So the form Eq.~(\ref{eq:cbarefact}) of the resummation has the
advantage of emphasizing the common RG origin of all soft
contributions. On the other hand, the form
Eq.~(\ref{eq:rescfdydisabd}) has the advantage of expressing the
resummed result as the exponentiation of the Mellin transform of
contributions that can be put in one-to-one correspondence to a
fixed order calculation.

\subsection{Two-scale resummation and the transverse momentum distribution}
\label{sec:tsr}

The multiscale case of prompt photon production considered in
Ref.~\cite{Bolzoni:2005xn}, or of transverse momentum distribution
discussed here, is a generalization of the approach of
Sect.~\ref{sec:rgres}, in which the process now depends 
on two different hard scales $Q_1$ and $Q_2$ and  in the soft
limit the phase space depends on two different dimensional variables of the form
Eq.~(\ref{eq:gen}). As discussed in  Sect.~\ref{sec:kin}
for a transverse momentum
distribution the two hard scales can be chosen as any pair out of
$Q^2$, $Q\pt$, $ \pt^2$. Because, as seen in Sect.~\ref{sec:phsp} the
two scales Eqs.~(\ref{eq:softscale}-\ref{eq:collscale}) are
respectively proportional to $Q^2$ and $Q\pt$ it is natural to pick
these as hard scales.

Assuming again full factorization, now also with
respect to the dependence on the two different scales~\cite{Bonciani:2003nt}, the coefficient
function takes the form
\begin{equation}\label{eq:cbarefactpt}
  C^{(0)}(N, Q^2_1,Q_2^2,\alpha_0,\epsilon)=C^{(0,\,c)}(Q^2_1,Q_2^2,\alpha_0,\epsilon) C^{(0,\,l_1)}(\bar\Lambda_{a_1}^2(N,\lambda_1^2),\alpha_0,\epsilon)C^{(0,\,l_2)}(\bar\Lambda_{a_2}^2(N,\lambda_2^2),\alpha_0,\epsilon),
\end{equation}
where the scales $\lambda_i$ are generally functions of the two scales
$Q_i$: $\lambda_i=\lambda_i(N,Q_1^2,Q_2^2)$.

A rerun of the same argument now leads to the
physical anomalous dimension
\begin{align}\label{eq:gampbardgpt}
\gamma(N, 1,Q_2^2/Q_1^2,\as(Q_2^2),\epsilon)&=
\bar g_0(\alpha(Q_1^2),Q_2^2)  
\nonumber\\&
+\int_{Q_1^2}^{\bar\Lambda_{a_1}^2(N,\lambda_1^2)} \frac{d\mu^2}{\mu^2}\bar g_1(\as(\mu^2))+\int_{Q_1^2}^{\bar\Lambda_{a_2}^2(N,\lambda_2^2)} \frac{d\mu^2}{\mu^2}\bar g_2(\as(\mu^2)).
\end{align}
Note that when solving the renormalization-group equation
we have  chosen one of the two hard scales,
namely $Q_1$, both as renormalization and factorization scale; of
course nothing prevents one from re-expressing the result for generic choices
of these scales.
We thus end up with a resummed coefficient function of the form
\begin{align}\label{eq:rescf}
&  C(N,Q^2_1/\mu^2,Q^2_2/\mu^2,\as(\mu^2))=C^{(c)}\left(\as(Q^2),\frac{Q^2}{\mu^2}\right)\nonumber\\&\qquad\times \exp\left[
  \int_1^{N^a_1}\frac{dn_1}{n_1} \int_{n_1 k^2}^{\lambda_1^2}
  \frac{dk_1^2}{k_1^2}\bar g_1(\as(k_1^2))+
  \int_1^{N^a_2}\frac{dn_2}{n_2} \int_{n_2 k^2}^{\lambda_2^2}
  \frac{dk_2^2}{k_2^2}\bar g_2(\as(k^2))\right],
\end{align}
where the functions $\bar g_i$ are power series in $\as$ of the
form of Eq.~(\ref{eq:gfun}). 

We come now to the case of the DY or Higgs transverse momentum distributions,
discussed in the previous section. In this case, the leading-order
process always has a parton in the final state, and we must thus
distinguish between two partonic subchannels, according to the species
(quark or gluon) of final-state parton. The final-state parton is necessarily of the same species of one of the initial-state partons.
The other initial-state parton is
a quark for the DY process, and a gluon for Higgs.  The argument of
Sect.~\ref{sec:phsp} then shows that the phase space factors into a
phase space with DY-like kinematics, characterized by the scale
$\Lambda^2_{\rm DY}$ Eq.~(\ref{eq:softscale}), and  a
phase space with DIS-like kinematics, characterized by the scale
$\Lambda^2_{\rm DIS}$ Eq.~(\ref{eq:collscale}). Assuming factorization
of the full amplitude in the soft limit the coefficient  function
factorizes according to Eq.~(\ref{eq:cbarefactpt}). We thus arrive at
the resummed expression
\begin{align}\label{eq:rescfpt}
&  C_{ij}(N,Q^2/\mu^2, Q\pt/\mu^2,\as(\mu^2))=C^{(c)}\left(\as(Q^2),\frac{Q^2}{\mu^2}\right)
\nonumber\\
&\qquad\times \exp\left[
  \int_1^{N^2}\frac{dn_1}{n_1} \int_{n_1 \mu^2}^{Q^2}
  \frac{dk_1^2}{k_1^2}\bar g^{(i)}_1(\as(k_1^2))+
 \int_1^{N}\frac{dn_2}{n_2} \int_{n_2 \mu^2}^{Q \pt}
\frac{dk_2^2}{k_2^2}\bar g^{(j)}_2(\as(k_2^2))\right],
\end{align}
where $Q^2$ is given by Eq.~(\ref{q2}), and $j$ denotes the species of
outgoing parton in the Born process, and $i$ is a quark for DY and a
gluon for Higgs.

Using further the expression Eq.~(\ref{eq:rescfdydisabd}) of the
resummed DY and DIS coefficient functions we get
\begin{align}
&C_{ij}(N,Q^2/\mu^2, Q\pt/\mu^2,\as(\mu^2))=g_0^{ij}\left(\as(Q^2),\frac{Q^2}{\mu^2}\right)
\nonumber\\
&\quad \times \exp\Bigg\{\int_0^1dx\,\frac{x^{N-1}-1}{1-x}
\[D^i[\as(Q^2(1-x)^2)]\
 +\int_{\mu^2}^{Q^2(1-x)^2}\frac{dk^2}{k^2}A^i[\as(k^2]\]
\nonumber\\
 &\qquad\qquad
 +\int_0^1dx\,\frac{x^{N-1}-1}{1-x} \[B^j[Q \pt(1-x)]
+ \int_{\mu^2}^{Q \pt (1-x)}\frac{dk^2}{k^2}A^j[\as(k^2)]\]\Bigg\}.
 \label{eq:rescfptabd}
\end{align}
Note that the function $B$ that characterizes radiation from the
outgoing parton now also carries an index according to whether this is
a quark or gluon: so it corresponds for instance to either standard
DIS (quark) or deep-inelastic
Higgs production in photon-gluon fusion (gluon)~\cite{Moch:2005ba}.

Equations~(\ref{eq:rescfpt}-\ref{eq:rescfptabd}) are the main result
of this paper and are the desired all-order generalization of the NLL
resummation of Ref.~\cite{deFlorian:2005fzc}. In the next section we
will compare them directly to the known NLO result and verify that
they agree with it, which may not be immediately obvious.

\section{Resummed results}
\label{sec:resexp}
Explicit resummed expressions can be obtained by performing the
integrals in Eqs.~(\ref{eq:rescfptabd}-\ref{eq:rescfpt}), in terms of
the coefficients of the expansion of the QCD $\beta$ function and of 
the functions $A_i$, $B_i$, $D_i$ 
(Eq.~(\ref{eq:rescfptabd})) or $g_i$ (Eq.~(\ref{eq:rescfpt})) in
powers of $\as$. These expressions were given up to NLL in
Ref.~\cite{deFlorian:2005fzc}, and are given at NNLL in the
Appendix. Below we check that our results agree with previous known
resummed or fixed-order results.

\subsection{Next-to-leading resummation}
\label{sec:nll}
As repeatedly mentioned, threshold resummation of transverse momentum
distributions up to NLL accuracy was given in
Ref.~\cite{deFlorian:2005fzc}. In that reference,  the resummed
expression of the coefficient function is written in the form 
\be\label{eq:defl}
 C_{ij}(N,Q^2/\mu^2, Q\pt/\mu^2,\as(\mu^2))=g_0^{ij}(\as(Q^2)) \Delta_i (Q\pt)\Delta_j (Q\pt)J_k(Q\pt)\Delta^{\rm int}_{ijk}(Q\pt)
\ee
where $k$ denotes the Born-level outgoing parton and
\begin{align}
&\ln\Delta_i(Q\pt)=\int_0^1dx\,\frac{x^{N-1}-1}{1-x}\int_{\mu^2}^{Q\pt(1-x)^2}\frac{dq^2}{q^2}\,A^i(\as(q^2))
\\
&\ln J_k(Q\pt)=\int_0^1dx\,\frac{x^{N-1}-1}{1-x}\[\int_{Q\pt(1-x)^2}^{Q\pt(1-x)}\frac{dq^2}{q^2}\,A^k(\as(q^2))+B^k(\as(Q\pt(1-x)))\]
\\
&\ln\Delta_{\rm int}^{ijk}(Q\pt)=\int_0^1dx\,\frac{x^{N-1}-1}{1-x}D^{ijk}(\as(Q\pt(1-x)^2)),
\end{align}
where the functions $A^i$ and $B^i$ are as in
Eq.~(\ref{eq:rescfptabd}) and
\be
D_1^{ijk}=-(A_1^i+A_1^j-A_1^k)\ln\frac{\pt}{Q},
\label{D1}
\ee
with $A_{1\,i}$ the first-order coefficient in the expansion of the
function $A^i(\as)$ in powers of its argument.
Also as mentioned, what is called $Q^2$ in Ref.~\cite{deFlorian:2005fzc} 
corresponds to what we call $Q\pt$ here.

In order to compare with our resummed expression Eq.~(\ref{eq:rescfptabd}), note that 
\be
\label{eq:nllscale}
  \as(Q\pt(1-x)^2)\ln\frac{\pt}{Q}=\int_{Q^2(1-x)^2}^{Q\pt(1-x)^2}\frac{dq^2}{q^2}\,\as(q^2)
  +O(\as^2(Q^2)).
\ee
It follows that, to NLL accuracy, we can reabsorb the $\Delta_{\rm
  int}^{ijk}(Q\pt)$ term into a change of scale of the  $\Delta_i
(Q\pt)$ and $J_k(Q\pt)$ functions. Namely, Eq.~(\ref{eq:defl}) can be
rewritten in the rather simpler form
\be\label{eq:defln}
 C_{ij}(N,Q^2/\mu^2, Q\pt/\mu^2,\as(\mu^2))=g_0^{ij}(\as(Q^2))
 \Delta_i (Q^2)\Delta_j (Q^2)\bar J_k(Q^2,Q\pt),
\ee
where
\begin{equation}\label{eq:jbar}
\ln
\bar J_k(Q^2,Q\pt)=\int_0^1dx\,\frac{x^{N-1}-1}{1-x}\[\int_{Q^2(1-x)^2}^{Q\pt(1-x)}\frac{dq^2}{q^2}\,A^k(\as(q^2))+B^k[\as(Q\pt(1-x))]\].
\end{equation}
Exploiting the fact that, as mentioned,  the outgoing parton $k$ is
always equal to  either
$i$ or to $j$, and  using the same convention as in 
Eqs.~(\ref{eq:rescf}-\ref{eq:rescfptabd}), i.e. assuming that the
outgoing parton is $j$ we then get
\be\label{eq:deltajbar}
\ln\Delta_j(Q^2)+\ln\bar J_j(Q^2,Q\pt)=
\int_0^1dx\,\frac{x^{N-1}-1}{1-x}\[\int_{\mu^2}^{Q\pt(1-x)}\frac{dq^2}{q^2}\,A^j(\as(q^2))+B^j[\as(Q\pt(1-x))]\].
\ee
Recalling that the function $D$ in Eq.~(\ref{eq:rescfptabd}) starts at
$O(\as^2)$, at the NLL level Eq.~(\ref{eq:defln}) is seen to
coincide with Eq.~(\ref{eq:rescfptabd}), with  $\Delta_i (Q^2)$ identified with the
Drell-Yan like term, and $\ln\Delta_j(Q^2)+\ln\bar J_j(Q^2,Q\pt)$
Eq.~(\ref{eq:deltajbar}) identified with the DIS-like term. 

It is interesting to observe that the renormalization-group argument
makes clear that to all orders the dependence on $\pt$ is only through
the argument of $\as$ evaluated at the scale $\Lambda^2_{\rm DIS}$
Eq.~(\ref{eq:collscale}). Note however that it is always possible even
at the NNLL level and beyond to rewrite the resummed expression
Eq.~(\ref{eq:rescfptabd}) in the form of Ref.\cite{deFlorian:2005fzc},
Eq.~(\ref{eq:defl}). Indeed, this simply corresponds to
choosing  $Q\pt(1-x)^2$ instead of $Q^2(1-x)^2$ as upper limit of integration
in Eq.~(\ref{eq:rescfptabd}). This then produces and extra
contribution of the form of the right-hand side of
Eq.~(\ref{eq:nllscale}), but now in general with higher order powers
of $\alpha_s(q^2)$ under the integral. The integral can be performed
and expressed in terms of $\ln\frac{\pt}{Q}$ and
$\alpha_s(Q\pt(1-x)^2)$, thereby leading to higher order contributions
to $\ln\Delta_{\rm int}^{ijk}(Q\pt)$.

\subsection{Comparison to fixed-order calculation}
\label{sec:compfo}

As discussed  in the introduction, in Ref.~\cite{deFlorian:2005fzc} the expansion of the
resummed result up to NLO was compared numerically to the
fixed order result $O(\alpha_s^2)$   for Higgs production
in gluon fusion of Ref.~\cite{Glosser:2002gm}. At $O(\alpha_s^2)$,
because the LO is $O(\alpha_s)$, the coefficient function Eq.~(\ref{eq:rescfdydisabd})
 must be expanded up to  $O(\alpha_s)$, and it then contains a LL
$\ln^2N$ term, and, at NLL,  $\ln N$ and constant contributions. The
 constant is fixed by matching to the fixed-order calculation, and it
 determines the contribution to $g_0$ in Eq.~(\ref{eq:rescfdydisabd})
 that is linear in $\alpha_s$.

 In a pure RG
 approach, in which only the functional form of the resummation
 Eq.~(\ref{eq:rescfpt}) or  Eq.~(\ref{eq:rescfptabd})  is known,
 the coefficients of the NLL logarithmic terms are also fixed by
 comparison to the fixed NLO, and then used to predict results at NNLO
 and beyond. However, if one instead uses the form
 Eq.~(\ref{eq:rescfdydisabd}) of the resummed result, all the
 logarithmic coefficients up to NLL accuracy (so all logarithmic
 coefficients at NLO)  are predicted in terms of the
 universal, process-independent functions $A$ and $B$.

We thus get a nontrivial check by comparing the coefficients of the
logarithmically enhanced terms of the fixed-order result to those
obtained by  using the form
 Eq.~\eqref{eq:rescfptabd} of the resummation, truncated to NLL, and
 expanded to first order in $\alpha_s$. We perform this check in the
 gluon-gluon channel, where we have
\begin{align}
	&C_{gg}(N,Q^2/\mu^2, Q\pt/\mu^2,\as(\mu^2))=1 
	+ \frac{\alpha_s}{2\pi} \Bigg[ 
			\int_0^1dx\,\frac{x^{N-1}-1}{1-x} 
				 \int_{\mu^2}^{Q^2(1-x)^2}\frac{dk^2}{k^2}A^g_1
	\nonumber\\
	&\qquad\qquad
			+\int_0^1dx\,\frac{x^{N-1}-1}{1-x} 
			\(
				B_1^g + \int_{\mu^2}^{Q \pt (1-x)}\frac{dk^2}{k^2}A^g_1
			\)\Bigg]
			+O\(\frac{\alpha_s}{2\pi}\)^2+O(\ln^0 N),
\end{align}
having noticed that the $D$ function starts at $O(\alpha_s^2)$,
that the contribution proportional to $A_2$, even though is
the same order as $B_1$, also starts contributing at $O(\alpha_s^2)$,
and finally that the prefactor function $g_0$ only starts contributing to
logarithmically enhanced terms at order $\alpha_s^2$.

The $k^2$ integrals are trivial to perform, and the $x$ integrals can be
obtained by differentiating the generating function
\begin{equation}
	J_\varepsilon = \int_{0}^{1}dx\, \frac{x^{k-1}}{(1-x)^{1-\epsilon}} = \frac{\Gamma(k)\Gamma(\epsilon)}{\Gamma(k+\epsilon)},
\end{equation}
see e.g.\ Ref.~\cite{Forte:2002ni}.
We then get
\begin{equation}
	C_{gg}(N,Q^2/\mu^2, Q\pt/\mu^2,\as(\mu^2))=1+\frac{\as}{2\pi}
\[	c_2 \ln^2N+
	c_1(\pt^2) 	\ln N
	+ O\(\ln^0 N\)\]+O(\as^2)
\end{equation}
with
\begin{align}\label{eq:c2rexp}
	c_2 =& 3 A_1^g
	\\\label{eq:c1rexp}
	c_1(\pt^2) =&2A_1^g\( 3 \gamma+2 \ln\frac{\mu^2}{Q^2}
	-\ln\frac{\pt}{Q}\)-2B_1^g.
\end{align}

Extracting the corresponding coefficients from the fixed-order NLO
result of Ref.~\cite{Glosser:2002gm} is not entirely trivial, because
the result given there is fully differential, so it must be integrated
in rapidity in order to get the transverse momentum distribution. The
computation is presented  in Appendix~\ref{sec:trmom}. The final
result is given in  Eqs.~(\ref{NLO_pt_c2}-\ref{NLO_pt_c1}), and it
matches the expansion of the resummed result Eqs.~(\ref{eq:c2rexp}-\ref{eq:c1rexp}).

\section{Outlook}
\label{sec:conc}
In this paper we have derived threshold resummation for transverse
momentum distributions by using a renormalization group approach to
resummation originally developed in
Refs.~\cite{Contopanagos:1996nh,Forte:2002ni}, extended to prompt
photon production in Ref.~\cite{Bolzoni:2005xn}, and here to the more
general case of processes with two soft scales, of which prompt photon production
and transverse momentum distributions are specific instances. From a
technical point of view, the main difference to previous work is that
a transverse momentum distribution not only has two different soft
scales in the threshold limit, but also  two distinct hard scales, which
makes the phase-space analysis that is at the heart of the
renormalization group argument somewhat more involved.

Our result generalizes to all orders previous NLL results of
Ref.~\cite{deFlorian:2005fzc}. In comparison to them, the
renormalization group approach leads to a somewhat more compact
resummed expression, also at NLL. In paricular, there is no need to
introduce a
process-dependent interference contribution specific  of
transverse momentum distribution~\cite{deFlorian:2005fzc}: this  is reabsorbed by a suitable
choice of scale. Rather, the result is entirely expressed in terms of
the resummation for the Drell-Yan-like and DIS-like structures in which
it is decomposed.

The resummed results presented here  may be used,  matched to fixed
order calculations, in order to accelerate their perturbative
convergence, and also, for the construction of approximations to
unknown higher fixed-order results~\cite{Ball:2013bra}. On a more theoretical note, these results can be useful
in the construction of joint resummed results, in which soft
resummation is combined with high-energy resummation~\cite{Bonvini:2018ixe} or with
transverse-momentum resummation~\cite{Muselli:2017bad}. Specifically,
they could shed light on the common origin of soft and
transverse-momentum resummed logarithms, thereby leading to more
powerful joint resummed expression. This is the subject of ongoing
investigations and it will be left to future work.

\section*{Acknowledgments}

SF thanks Kirill Kudashkin and Tanjona Rabemananjara for several discussions.
SF is  supported by the European Research Council under
the European Union's Horizon 2020 research and innovation Programme
(grant agreement n.740006). The work of GR is supported by the Italian Ministry of Research (MIUR) under grant PRIN 20172LNEEZ.

\appendix
\section{Explicit resummed results}
\label{sec:expres}
The exponent of a resummed coefficient function is conveniently
written as a power expansion in $\as(Q^2)$ at fixed 
$\as(Q^2)\ln N$. To this purpose, we define a resummation variable 
\be
\lambda=b_0\as(Q^2)\ln N
\ee
where $b_0$ is the first coefficient in the expansion of the QCD $\beta$ function:
\begin{align}
\mu^2\frac{d\as(\mu^2)}{d\mu^2}&=-b_0\as^2(\mu^2)-b_1\as^3(\mu^2)-b_2\as^4(\mu^2)+O(\as^5);\\
b_0&=\frac{11 C_A - 4 T_f n_f}{12\pi},\label{b_0}
\\
b_1&=\frac{17 C_A^2 - 10C_A n_f T_f - 6 C_F n_f T_f)}{24\pi^2},
\\
b_2&=\frac{1}{128\pi^3}\(2857 - \frac{5033}{9}n_f + \frac{325}{27}n_f^2\).
\end{align}

The resummed exponent, i.e. the logarithm of the resummed coefficient
function, is then written
as an expansion in powers of $\alpha_s$ at fixed $\lambda$, so that
including the first $k+1$ orders of this expansion, that we call
$g_k^{ij}\(\lambda,\frac{\pt}{Q}\)$, 
resummation at N$^k$LL level is obtained. 
Up to and including NNLL we thus get
\begin{align}
&C_{ij}(N,Q^2/\mu^2, Q\pt/\mu^2,\as(\mu^2))=g_0^{ij}(\as(Q^2))
\nonumber\\
&\qquad\qquad\exp\[\frac{1}{\as(Q^2)} g^{ij}_1\(\lambda,\frac{\pt}{Q}\)+g^{ij}_2\(\lambda,\frac{\pt}{Q}\)+\as(Q^2)g_3^{ij}\(\lambda,\frac{\pt}{Q}\)+O(\alpha_s^3)\].
\label{eq:CNNLL}
\end{align}
Note that the expansion of $g_1$ in powers of $\alpha_s$ starts at
order $\alpha_s^2$ while the expansion of all $g_i$ with $i>1$ starts
at $O(\alpha_s)$. 
In order to achieve N$^k$LL accuracy of the coefficient function, it is
necessary also to include the first $k+1$ orders in the expansion
of the prefactor $g_0^{ij}(\as(Q^2))$ in powers of $\alpha_s$. At
N$^k$LL order the coefficients of all terms proportional to
$\alpha_s^n \ln^{2n-2k} N$ are then correctly predicted.

The coefficients $g_k^{ij}\(\lambda,\frac{\pt}{Q}\)$ are given in terms of the functions $A^i(\as),B^i(\as),D^i(\as)$ defined in Eq.~(\ref{eq:rescfdydisabd}), which in turn have perturbative expansions
\begin{align}\label{eq:aexp}
A^i(\as)&=\sum_{n=1}^\infty A_n^i\(\frac{\as}{\pi}\)^n
\\\label{eq:bexp}
B^i(\as)&=\sum_{n=1}^\infty B_n^i\(\frac{\as}{\pi}\)^n
\\\label{eq:dexp}
D^i(\as)&=\sum_{n=2}^\infty D_n^i\(\frac{\as}{\pi}\)^n.
\end{align}
The universal coefficients $A^i$ and $B^i$ and the 
coefficients $D^i$ for Drell-Yan and Higgs production are given up to
N$^3$LL in Ref.~\cite{Moch:2005ba}.

Substituting the expansions Eqs.~(\ref{eq:aexp}-\ref{eq:dexp}) in the
resummed expression Eq.~(\ref{eq:rescfdydisabd}) we get, up to NNLL,
\begin{align}\label{eq:g1r}
g_1^{ij}\(\lambda,\frac{\pt}{Q}\)&=\frac{1}{2 b_0^2 \pi}\[A_1^i \(2\lambda + (1 - 2\lambda)\ln(1 - 2\lambda)\) + 
 2 A_1^j \(\lambda+ (1 - \lambda) \ln(1 -\lambda)\)\]\end{align}
\begin{align}
 \label{eq:g2r}
g_2^{ij}\(\lambda,\frac{\pt}{Q}\)&=
\frac{A_1^i}{4 b_0^3\pi}
\[4\lambda \(b_1 + b_0^2\ln\frac{\mu^2}{Q^2}\) + 
\ln(1 - 2 \lambda) (2 b_1 - 4 b_0^2 \gamma + 
    b_1 \ln(1 - 2\lambda))
\]
\nonumber\\
&-\frac{A_2^i}{2 b_0^2 \pi^2}\[2\lambda +\ln(1 - 2\lambda)\]
\nonumber\\
&+\frac{A_1^j}{2 b_0^3\pi}\[2\lambda (b_1 + b_0^2\ln\frac{\mu^2}{Q^2}) + 
 2(b_1 + b_0^2 (\ln\frac{\pt}{Q}-\gamma))\ln(1 - \lambda)+ 
b_1\ln^2(1 -\lambda)\]
\nonumber\\
&-\frac{A_2^j}{b_0^2\pi^2}\[\lambda+\ln(1-\lambda)\]
+\frac{B_1^j}{b_0\pi}\ln(1-\lambda).
\end{align}

\begin{align}\label{eq:g3r}
&g_3^{ij}\(\lambda,\frac{\pt}{Q}\)
=
\frac{1}{12 b_0^4 \pi^3}
\Bigg\{\frac{1}{2 \lambda-1}
  \Big[6 b_0 (
  2 b_0 \lambda (-\lambda A_3^i + b_0 \pi D_2^i) 
  \nonumber\\
&+\pi A_2^i(2 b_1 \lambda (1 + \lambda) 
 - 2 b_0^2 \lambda (2 \gamma + \ln\frac{\mu^2}{Q^2} - 2 \lambda \ln\frac{\mu^2}{Q^2}) + 
          b_1 \ln(1 - 2 \lambda)))
\nonumber\\
&+ \pi^2 A_1^i (-2 \lambda (-12 b_0^2 b_1 \gamma - 6 b_0 b_2 (\lambda-1) + 
          6 b_1^2 \lambda 
+b_0^4 (12 \gamma^2 + (-3 + 6 \lambda) \ln^2\frac{\mu^2}{Q^2} + 2 \pi^2)) 
\nonumber\\
&- 
       3 \ln(1 - 2 \lambda) (2 b_0 b_2 - 4 b_0^2 b_1 \gamma + 
          4 b_1^2 \lambda - 4 b_0 b_2 \lambda + b_1^2 \ln(1 - 2 \lambda)))\Big]
\nonumber\\
&+\frac{1}{\lambda-1}\Big[6 b_0 (\lambda (b_1 (2 + \lambda) - 
          2 b_0^2 \(\gamma- \ln\frac{\pt}{Q} + (1-\lambda)\ln\frac{\mu^2}{Q^2}  \)) \pi A_2^j
\nonumber\\
&          + 
       b_0 \lambda (-\lambda A_3^j 
+2 \pi (-(b_1 - b_0^2 \(\gamma -\ln\frac{\pt}{Q}\)) \pi B_1^j + b_0 B_2^j)) + 2 b_1 \pi (A_2^j - b_0 \pi B_1^j) \ln(1 - \lambda)) 
\nonumber\\
&+ \pi^2 A_1^j
       (\lambda (6 b_0 b_2 ( \lambda-2) - 6 b_1^2 \lambda + 
          12 b_0^2 b_1 \(\gamma - \ln\frac{\pt}{Q}\)
\nonumber\\
&- 
          b_0^4 (6 ((\lambda-1) \ln^2\frac{\mu^2}{Q^2} + 
          \(\gamma - \ln\frac{\pt}{Q}\)^2) + \pi^2)) - 
       6 \ln(1 - \lambda) (2 (b_1^2 \lambda + b_0 (b_2 - b_2 \lambda) 
       \nonumber\\
       &- 
             b_0^2 b_1 \(\gamma -\ln\frac{\pt}{Q}\)) + b_1^2 \ln(1 - \lambda)))\Big]
             \Bigg\}
\end{align}
In these expressions, $\mu^2$ is the factorization scale, while the renormalization scale $\mur^2$ was taken to be equal to $Q^2$. The dependence on the renormalization scale can easily be restored by shifting the argument of $\as$. To NNLL accuracy, this amounts to replacing
\be
\as(Q^2)=\as(\mur^2) + \as^2(\mur^2) b_0 \ln\frac{\mur^2}{Q^2} + \as^3(\mur^2) \(b_1\ln\frac{\mur^2}{Q^2}   + b_0^2 \ln^2\frac{\mur^2}{Q^2} \)
\ee
in Eq.~(\ref{eq:CNNLL}), and expanding in powers of $\as(\mur^2)$ at
\be
\lambda_{\sss R}=b_0\as(\mur^2)\ln N
\ee
fixed. As a consequence, the functions $g_2,g_3$ are modified as follows:
\begin{align}
g_2^{ij}\(\lambda,\frac{\pt}{Q}\)\to& g_2^{ij}\(\lambda_{\sss R},\frac{\pt}{Q}\)
-\frac{1}{2\pi b_0}\[A_1^i (2\lambda_{\sss R} +\ln(1 - 2\lambda_{\sss R}))+2A_1^j(\lambda_{\sss R} +\ln(1 - \lambda_{\sss R}))\]\ln\frac{\mur^2}{Q^2}
\label{eq:g2mur}
\\
g_3^{ij}\(\lambda,\frac{\pt}{Q}\)\to &g_3^{ij}\(\lambda_{\sss R},\frac{\pt}{Q}\)
+\frac{\lambda_{\sss R}^2 (2 (1 - \lambda_{\sss R}) A_1^i + (1 - 2 \lambda_{\sss R}) A_1^j)}
{2 \pi(1 - \lambda_{\sss R}) (1 - 2\lambda_{\sss R})}\ln^2\frac{\mur^2}{Q^2}
\nonumber\\
&+\frac{1}{2\pi^2 b_0^2 (1 - \lambda_{\sss R}) (1 - 2 \lambda_{\sss R})}
\Bigg[
\nonumber\\
&
-\pi A_1^i (1 - \lambda_{\sss R}) 
\(2 \lambda_{\sss R} (b_1 - b_0^2 (2 \gamma + \ln\frac{\mu^2}{Q^2} (1- 2 \lambda_{\sss R}))) 
+ b_1 \ln(1 - 2 \lambda_{\sss R})\) 
 \nonumber\\
&+2b_0 \lambda_{\sss R} \(2 (1 - \lambda_{\sss R}) \lambda_{\sss R} A_2^i + (1 - 2 \lambda_{\sss R}) (\lambda_{\sss R} A_2^j - b_0 \pi B_1^j)\) 
 \nonumber\\
 &
 - 2\pi A_1^j (1 - 2 \lambda_{\sss R}) \(\lambda_{\sss R} (b_1 - b_0^2 (\gamma + (1 - \lambda_{\sss R}) \ln\frac{\mu^2}{Q^2} 
 -\ln\frac{\pt}{Q}))
 + b_1 \ln(1 - \lambda_{\sss R})\)
\Bigg]\ln\frac{\mur^2}{Q^2}
\label{eq:g3mur}
\end{align}

Results up to NLO  were already given in Ref.~\cite{deFlorian:2005fzc}.
We have shown in Sect.~\ref{sec:nll} that our results are in agreement
with that reference, and we further checked explicitly that
Eqs.~(\ref{eq:g1r},\ref{eq:g2r},\ref{eq:g2mur}) agree with results presented there,
once account is taken for the different notation and choice of hard
scale.  The NNLL result Eqs.~(\ref{eq:g3r},\ref{eq:g3mur}) is a new result of this paper. 

\section{The NLO transverse momentum distribution}
\label{sec:trmom}

The NLO computation  of Ref.~\cite{Glosser:2002gm} determines the
double differential cross section in transverse momentum and rapidity
at the partonic level for Higgs production in gluon fusion.
The computation is performed in the heavy top mass limit; however, in
the soft limit the heavy top approximation is exact. In order to
compare this result to the resummed result we must  integrate in
rapidity the double differential cross section.

We concentrate on the gluon-gluon channel, for which the expansion of
the resummed result was given in Sect.~\ref{sec:compfo}.
In Ref.~\cite{Glosser:2002gm}
the the double differential partonic cross-section is written as 
\begin{equation}	\label{eq:glosser_sigma_pert_expansion}
	\frac{d\hat{\sigma}_{ij}}{d\pt^2 dy}	=
	\frac{\sigma_0}{\hat s} 									%sez LO inclusiva
	\left[
	\frac{\as(\mur^2)}{2\pi}	G_{gg}^{(1)}	%sez LO
	+
	\left(
	\frac{\as(\mur^2)}{2\pi}
	\right)^2	G_{gg}^{(2)}							%sez LO
	+O((\as)^3)
	\right],
\end{equation}
where $\sigma_0$ is the leading-order total cross-section, and the 
logarithmically enhanced (either in the soft or the small $\pt$ limit)
contribution to the NLO term  
$G_{gg}^{(2)}$ is given in Eq. (3.17) of Ref.~\cite{Glosser:2002gm}.

The rapidity integral can be
computed by performing a change of variable suggested in
Ref.~\cite{Ravindran:2002dc} from rapidity $y$ to $K^2$, the invariant
mass of the two  final state parton system. The  two variables are related by
\begin{equation}	\label{eq:y_as_fct_of_Q2}
	\sinh y =	
	\pm \frac{\sqrt{ (\hat s + m^2 - K^2)^2 - 4\hat s \Et^2}}
	{2 \sqrt{\hat s} \Et}.
\end{equation}
By a further change of variables $K^2 \rightarrow q=K^2/K^2_{\rm max}$,
where $K^2_{\rm max}$ is the upper bound of the $K^2$ integral, the
rapidity integrals,         up to terms of order $O(1-x)$,
 are all reduced to integrals of the general  form
\begin{equation}\label{eq:ik}
I_k=
\int_{0}^{1} 
	\left[
	\frac{\ln^k q}{q}
	\right]_+
	\frac{1}{\sqrt{(1 - q)}}
	dq.
\end{equation}
        This is done by starting with the plus distributions
        contained in $G_{gg}^{(2)}$ of Ref.~\cite{Glosser:2002gm},
        which are expressed in terms of the scaling variables $z_t$ and
        $z_u$ respectively given by
        \begin{align}
          z_t=\frac{-t}{K^2-t};\qquad z_u=\frac{-u}{K^2-u},
          \end{align}
and  exploiting the identities
        between plus distributions
\begin{align}
\frac{z_t}{-t} \(\frac{1}{1-z_t} \)_+ &=\frac{1}{K^2_{\rm max}}\left\lbrace
\[\frac{1}{q} \]_+ + \delta(q) \ln \frac{K^2_{\rm max}}{-t}\right\rbrace
\\
\frac{z_t}{-t} \( \frac{\ln(1-z_t)}{1-z_t} \)_+ &=\frac{1}{K^2_{\rm max}}\left\lbrace
\[ \frac{\ln(q)}{q} \]_++ \ln \frac{K^2_{\rm max} z_t}{-t} \[\frac{1}{q} \]_+ 
+\frac{\delta(q)}{2} \ln^2 \frac{K^2_{\rm max}}{-t}
\right\rbrace
\end{align}
and similarly for $z_u$.

        The integrals  $I_k$ Eq.~(\ref{eq:ik}) can be computed by differentiating with respect to $\epsilon$ the generating function
\begin{equation}
	I_{\epsilon} =
	\int_{0}^{1}
	\frac{1}{q^{-1+\epsilon}}	
	\left[
	\frac{1}{\sqrt{(1 - q)}}
	- 1
	\right]
	=
	\frac{\Gamma \(\frac{1}{2}\)	\Gamma(\epsilon)}
		{\Gamma \(\frac{1}{2} + \epsilon\)}
	- \frac{1}{\epsilon}.
\end{equation}
This leads to an expression of the  differential cross section that contains threshold logarithms of the form~\cite{Muselli:2017bad}
$	\left(
	\frac{\ln^k (1-x)}{\sqrt{(1-x)(1-ax)}}
	\right)_+$, where $a=\(\frac{\Et+\pt}{m}\)^2$,
whose Mellin transform  can be computed from the generating function
\begin{align} \label{eq:melsqrt}
	D_\eta^a(N) =
	\int_{0}^{1}
	\frac{x^N}
	{\sqrt{(1-x)(1-ax)}}
	(1-x)^\eta
	dx
	=
	\frac{\Gamma (N) \Gamma \( \eta + \frac{1}{2}\)}
		{\Gamma \(N + \eta + \frac{1}{2}\)}
	\prescript{}{2}{F}_1 \left(\frac{1}{2},
								N;
								N + \eta + \frac{1}{2};
								a^2 \right),
\end{align}
where $\prescript{}{2}{F}_1$ is a hypergeometric function.
When differentiating Eq.~(\ref{eq:melsqrt}) with respect to $\eta$ one
should keep in mind that for finite $\pt$ the logarithmic derivative
of the hypergeometric function is power suppressed in the large-$N$
region. 

The $\pt$ distribution at NLO in the threshold limit is finally found
to take the form
\begin{equation} \label{eq:NLO_cs_Mellin_with_c_coeffs}
		\frac{d\sigma ^{NLO}}{d\pt^2} (N,\pt^2) =	
		\frac{\alpha_s}{2\pi}				
		\frac{d\sigma ^{LO}}{d\pt^2} (N,\pt^2) 	
		 \left\lbrace
		c_2 \ln^2(N) +
		c_1(\pt^2) 	\ln(N) 	
			+ O\(\ln^0 N\)
		\right\rbrace		
\end{equation}
with
\begin{align}
	c_2 =& 3 N_c, 
	\label{NLO_pt_c2}
	\\
	c_1(\pt^2) =& 6 \gamma N_c-4 N_c \ln \left(\frac{Q^2}{\mu^2}\right)
	-2 N_c \ln \(\frac{\pt}{Q}\)+2\pi b_0,
	\label{NLO_pt_c1}
\end{align}
where $b_0$ is given in Eq.~(\ref{b_0}).
Eqs.~(\ref{NLO_pt_c2},\ref{NLO_pt_c1})
are seen to coincide with the values obtained by expanding the resummed coefficient function to order $\as$, 
Eqs.~(\ref{eq:c2rexp},\ref{eq:c1rexp}), since
\be
A_1^g=C_A=N_c;\qquad B_1^g=-\pi b_0.
\ee

\bibliographystyle{UTPstyle}
\bibliography{FRR_v2}

\end{document}